\documentclass{article}

\usepackage{arxiv}

\usepackage[utf8]{inputenc} % allow utf-8 input
\usepackage[T1]{fontenc}    % use 8-bit T1 fonts
\usepackage{hyperref}       % hyperlinks
\usepackage{url}            % simple URL typesetting
\usepackage{booktabs}       % professional-quality tables
\usepackage{amsfonts}       % blackboard math symbols
\usepackage{nicefrac}       % compact symbols for 1/2, etc.
\usepackage{microtype}      % microtypography
\usepackage{lipsum}
\usepackage{color}
\usepackage{amsmath}
\usepackage[
backend=biber,
style=numeric,
citestyle=numeric
]{biblatex}

\usepackage{graphicx}
\graphicspath{{./figures/}}
\usepackage{braket}

\newcommand{\mi}{\mathrm{i}}
\newcommand{\op}[1]{\hat{#1}}
\newcommand{\tr}{\mathrm{tr}}

\title{Intrinsic decoherence dynamics in the three-coupled harmonic oscillators interaction}

\author{
  Alejandro R. Urz\'ua and H\'ector M. Moya Cessa\\
  Coordinaci\'on de \'Optica,\\ Instituto Nacional de Astrof\'isica, \'Optica y Electr\'onica\\ Luis Enrique Erro \#1, Tonantzintla, Puebla, M\'exico
}

\begin{document}
\maketitle

\begin{abstract}
    Applying the Milburn equation to describe intrinsic decoherence, we study the interaction of three-coupled quantum harmonic oscillators or quantized fields. We give an explicit solution for the complete equation, {\it i.e.}, beyond the usual second order approximation used to arrive to the Lindblad form. Then we calculate the expectation value of the number operator of each oscillator or mode for one of the modes given in an initial coherent state. 
\end{abstract}

\section{Introduction}\label{sec:intro}
Decoherence is a topic of great interest in quantum mechanics as it avoids that non-classical states of a given quantum system from  maintaining their significant properties, particularly the system's purity, producing rapidly a mixture of states {\it i.e.}, in a fast fashion, erasing its non-classicality. In 1991, Milburn \cite{Milburn1991} proposed a  model for intrinsic decoherence in quantum mechanics based on a simple modification of the Schr\"odinger equation. He assumed that the system evolves by a random sequence of unitary phase changes on sufficiently small time scales and managed to produce a Lindblad equation where the Hamiltonian is the operator involved in the Master Equation. The usual evolution is recovered to first order in an expansion parameter related to the speed at which the coherences are lost. Moya-Cessa {\it et al.} \cite{MoyaCessa1993} showed that in the atom-field interaction the loss of coherences prevents the revivals to occur for the atomic population inversion. Mohamed {\it et al.} \cite{Mohamed2021} analyzed the robustness of quantum correlations of the nearest neighbour and the next-to-neighbour qubits in an intrinsic noise model describing the dynamics of the decoherence for a system formed by three-qubit Heisenberg $XY$ chain; Yang {\it et al.} \cite{Yang2017} determined the performance of quantum Fisher information of the two-qutrit isotropic Heisenberg $XY$ chain subject to decoherence. Muthuganesand and Chandrasekar \cite{Muthuganesan2021} applied intrinsic decoherence when studying an exactly solvable model of two interacting spin-$\frac{1}{2}$ qubits described by the Heisenberg anisotropic interaction. Zheng and Zhang \cite{Zheng2017} applied Milburn's scheme to study the entanglement in the Jaynes-Cummings model, where a pair of atoms undergo  Heisenberg type interactions; He and Chao \cite{He2021} studied the coherence dynamics of two atoms in a Kerr-like medium. Chlih \textit{et al.} \cite{Chlih2021} used the intrinsic decoherence to study a variety of initial states, where they obtain the temporal evolution of quantum correlations in a two-qubit XXZ Heisenberg spin chain model subject to a Dzyaloshinskii–Moriya (DM) interaction and  to an external uniform magnetic field. Guo-Hui and Bing Bing \cite{GuoHui2015} estimated the quantum discord of two qubits that loose coherence through  intrinsic mechanisms. Leon-Montiel \textit{et. al} \cite{LenMontiel2015} used the concept that off-diagonal terms induce decoherence when a disorder is established. Furthermore, Mohamed \textit{et. al} \cite{Mohamed2020} have used the intrinsic decoherence effect for two qubits interacting with a coherent field, with the purpose to protect the entropy and entanglement from the dipole-dipole interaction. Also, the intrinsic decoherence scheme offers and alternative to the study of coherence phenomena under symmetry-breaking, just as Gong \textit{et. al} \cite{Gong2018} showed that in a ring arrangement of coupled harmonic oscillators.  Decoherence without dissipation of a charged magneto-oscillator by using quantum non-demolition interactions in non-commutative phase-space has also been recently studied \cite{Germain}. 

In this work we propose to analyze the dynamics of photon population in a three-coupled harmonic oscillators when  intrinsic decoherence  takes place. Our aim is to estimate the decoherence of one of the oscillators, and the redistribution of the photon average number over the others.

The paper is organized as follows: in Sec. \ref{sec:intdeco} we solve analytically Milburn's intrinsic decoherence equation. However, we do not solve the approximated (Lindblad) equation but instead the complete Miburn equation, without any approximations, showing how averages may be performed for arbitrary operators; in Sec. \ref{sec:threef},  given the three-coupled harmonic oscillator Hamiltonian (under rotating wave approximation), which is nothing but the Hamiltonian of three interactin quantized fields, we show how diagonalization may be performed to, finally, obtain full analytical solutions for the photon number evolution when a selected initial condition is considered. In Sec. \ref{sec:concls} we give some remarks and conclusions.

%%%%%%%%%%%%%%%%%%%%%%%%%%%%%%%%%%%%%%%%%%%%%%%%%%%%%%%%%%%%%%%%%%%%%%%%%%%%%%
%%%%%%%%%%%%%%%%%%%%%%%%%%%%%%%%%%%%%%%%%%%%%%%%%%%%%%%%%%%%%%%%%%%%%%%%%%%%%%
\section{Intrinsic decoherence equation and solution}\label{sec:intdeco}
In 1993 Milburn \cite{Milburn1991} introduced a modified Scrh\"odinger equation to describe (intrinsic) decoherence in the form
\begin{equation}\label{0010}
    \dot{\rho}=\gamma \left(e^{-\mi\frac{\op{H}}{\gamma}}\rho e^{\mi\frac{\op{H}}{\gamma}}- \rho\right),
\end{equation}
where $\op{H}$ is the system's Hamiltonian, and $\gamma$ is the intrinsic decoherence constant, that in the original proposed model is regarded as a decreasing parameter of the decoherence, giving the attributes to determine the time scale of the coherence suppression. By expanding the exponentials above in Taylor series, and keeping terms up to second order, we obtain
\begin{equation}\label{0011}
    \dot{\rho}\approx\gamma \left(\left[1-\mi\frac{\op{H}}{\gamma}-\frac{\op{H}^{2}}{\gamma^2}\right]\rho \left[1+\mi\frac{\op{H}}{\gamma}-\frac{\op{H}^2}{\gamma^2}\right]- \rho\right),
\end{equation}
that can be written in the Lindblad form, namely,
\begin{equation*}
    \dot{\rho}=-\mi\left[\op{H},\rho\right]-\frac{1}{\gamma}\left[\op{H},\left[\op{H},\rho\right]\right],
\end{equation*}
where the Schr\"odinger equation is recovered when $\gamma\rightarrow\infty$. 

However, it is worth to notice that equation \eqref{0010} has the simple solution
\begin{equation}\label{0020}
    \rho(t)=e^{-\gamma t}e^{\op{S}t}\rho(0),
\end{equation}
where we have defined the superoperator
\begin{equation*}
    \op{S}\rho=\gamma e^{-\mi\frac{\op{H}}{\gamma}}\rho e^{\mi\frac{\op{H}}{\gamma}},
\end{equation*}
such that
\begin{equation*}
    e^{\op{S}t}\rho(0)=\sum_{k=0}^{\infty}\frac{(\gamma t)^k}{k!}\, \rho_{k},
\end{equation*}
with the $k$-th element of the density matrix $\rho$ defined by
\begin{equation}\label{expt}
    \rho_{k}=\ket{\psi_{k}}\bra{\psi_k}, \qquad \ket{\psi_{k}}=e^{-\mi k\frac{\op{H}}{\gamma}}\ket{\psi(0)},
\end{equation}
where $\ket{\psi(0)}$ is the system's wavefunction at $t = 0$.

Once the solution for $\rho(t)$ is obtained, we may find the expectation values for the observables $\op{A}$ as
\begin{equation*}
     \braket{\op{A}}  = \tr\{\rho(t)\op{A}\} = \tr\{\op{A} e^{-\gamma t}e^{\op{S}t}\rho(0)\},
\end{equation*}
that finally renders the complete time evolution of the observable quantities
\begin{equation}\label{average}
    \braket{\op{A}}  = e^{-\gamma t}\sum_{k=0}^{\infty}\frac{ (\gamma t)^k}{k!} \braket{\psi_{k} |\op{A}| \psi_{k}}.
\end{equation}

%%%%%%%%%%%%%%%%%%%%%%%%%%%%%%%%%%%%%%%%%%%%%%%%%%%%%%%%%%%%%%%%%%%%%%%%%%%%%%
%%%%%%%%%%%%%%%%%%%%%%%%%%%%%%%%%%%%%%%%%%%%%%%%%%%%%%%%%%%%%%%%%%%%%%%%%%%%%%
\section{Three interacting fields}\label{sec:threef}
For three-coupled time-independent quantum harmonic oscillators, we may write the Hamiltonian
\begin{equation}\label{H3}
\begin{aligned}
    \op{H} &= \omega\left(\op{a}_{1}^{\dagger}\op{a}_{1} + \op{a}_{2}^{\dagger}\op{a}_{2} + \op{a}_{3}^{\dagger}\op{a}_{3}\right) + \lambda\left(\op{a}_{1}^{\dagger}\op{a}_{2} + \op{a}_{2}^{\dagger}\op{a}_{1}\right)\\
    &+ g\left[\op{a}_{3}\left(\op{a}_{1}^{\dagger}+\op{a}_{2}^{\dagger}\right)+\op{a}_{3}^{\dagger}(\op{a}_{1}+\op{a}_{2})\right],
\end{aligned}
\end{equation}
where $\omega$ is the individual angular frequency of each oscillator, $\lambda$ and $g$ are the strength coupling constants between the oscillators. 
\subsection{Diagonalization}\label{sec:diag}
Next, we can start diagonalizing \eqref{H3}, by applying a rotation with the help of the transformation operator $\op{R}_{12}=e^{\theta\left(\op{a}_{2}^{\dagger}\op{a}_{1}-\op{a}_{1}^{\dagger}\op{a}_2\right)}$ \cite{Urzua2019}, whose action on the oscillator annihilation operators $\op{a}_{1}$ and $\op{a}_{2}$ are
\begin{align*}
    \op{R}_{12}\, \op{a}_{1}\, \op{R}_{12}^{\dagger} &= \op{a}_{1}\cos\theta + \op{a}_{2}\sin\theta, \\
    \op{R}_{12}\, \op{a}_{2}\, \op{R}_{12}^{\dagger} &= \op{a}_{2}\cos\theta - \op{a}_{1}\sin\theta.
\end{align*}
Applying this pair of transformations to the  Hamiltonian \eqref{H3}, and simplifying terms, we arrive to the new Hamiltonian, $\op{H}_{1}$, that has explicit dependence on $\theta$ 
\begin{align*}
    \op{H}_{1} &= \op{R}_{12}\, H\, \op{R}_{12}^{\dagger}\\ 
    &= \omega \op{a}_{3}^{\dagger}\op{a}_{3} + \omega\left(\op{a}_{1}^{\dagger}\cos\theta + \op{a}_{2}^{\dagger}\sin\theta\right)\left(\op{a}_{1}\cos\theta + \op{a}_{2}\sin\theta\right)\\
    &+ \omega\left(\op{a}_{2}^{\dagger}\cos\theta -  \op{a}_{1}^{\dagger}\sin\theta\right)\left(\op{a}_2\cos\theta - \op{a}_1\sin\theta\right)\\
    &+ \lambda\left[\left(\op{a}_{1}^{\dagger}\cos\theta + \op{a}_{2}^{\dagger}\sin\theta\right)\left(\op{a}_{2}\cos\theta - \op{a}_{1}\sin\theta\right)\right.\\
    &+ \left.\left(\op{a}_2^{\dagger}\cos\theta - \op{a}_{1}^{\dagger}\sin\theta\right)\left(\op{a}_{1}\cos\theta + \op{a}_{2}\sin\theta\right)\right]\\
    &+ g\left[\op{a}_{3}\left(\op{a}_{1}^{\dagger}\cos\theta + \op{a}_{2}^{\dagger}\sin\theta + \op{a}_{2}^{\dagger}\cos\theta - \op{a}_{1}^{\dagger}\sin\theta\right)\right.\\
    &+ \left. \op{a}_{3}^{\dagger}\left(\op{a}_{1}\cos\theta + \op{a}_{2}\sin\theta + \op{a}_{2}\cos\theta - \op{a}_{1}\sin\theta\right)\right],
\end{align*}
where we can get rid of the terms that vanishes at $\theta=\pi/4$, giving us the simpler Hamiltonian
\begin{align*}
    \op{H}_{1} &= \omega \op{a}_{3}^{\dagger}\op{a}_{3} + \frac{\omega}{2}\left(\op{a}_{1}^{\dagger}\op{a}_{1} + \op{a}_{2}^{\dagger}\op{a}_{2} + \op{a}_{1}^{\dagger}\op{a}_{2} + \op{a}_{2}^{\dagger}\op{a}_{1}\right)\\
    &+ \frac{\omega}{2}\left(\op{a}_{1}^{\dagger}\op{a}_{1} + \op{a}_{2}^{\dagger}\op{a}_{2} - \op{a}_{1}^{\dagger}\op{a}_{2} - \op{a}_{2}^{\dagger}\op{a}_{1}\right)\\ 
    &+ \frac{\lambda}{2} \left[\left(\op{a}_{1}^{\dagger} + \op{a}_{2}^{\dagger}\right)\left(\op{a}_{2} - \op{a}_{1}\right) + \left(\op{a}_{2}^{\dagger} - \op{a}_{1}^{\dagger}\right)\left(\op{a}_{1} + \op{a}_{2}\right)\right]\\
    &+ \frac{g}{\sqrt{2}}\left[\op{a}_{3}\left(\op{a}_{1}^{\dagger} + \op{a}_{2}^{\dagger} + \op{a}_{2}^{\dagger} - \op{a}_{1}^{\dagger}\right) + \op{a}_{3}^{\dagger}\left(\op{a}_{1}+\op{a}_{2}+\op{a}_{2}-\op{a}_{1}\right)\right],
\end{align*}
that may be rewritten as
\begin{equation}\label{H1}
    \op{H}_{1} = \omega_{-}\op{a}_{1}^{\dagger}\op{a}_{1} + \omega_{+}\op{a}_{2}^{\dagger}\op{a}_{2} + \omega \op{a}_{3}^{\dagger}\op{a}_{3}  + {\sqrt{2}} g\left(\op{a}_{3}\op{a}_{2}^{\dagger} + \op{a}_{3}^{\dagger}\op{a}_{2}\right),
\end{equation}
where we have defined the coefficients $\omega_{\pm} = \omega \pm \lambda$.

A second transformation, $\op{R}_{2}=e^{\phi\left(\op{a}_{2}^{\dagger}\op{a}_{3} - \op{a}_{3}^{\dagger}\op{a}_2\right)}$, can be performed on the Hamiltonian given in equation \eqref{H1}, who may eliminate the remaining interaction terms if we choose adequately $\phi$, so we arrive to
\begin{align*}
    \op{H}_{2} &= \op{R}_{2}\, \op{H}_{1}\, \op{R}_{2}^{\dagger}\\ 
    &= \omega\left(\op{a}_{3}^{\dagger}\cos\phi + \op{a}_{2}^{\dagger}\sin\phi\right)\left(\op{a}_{3}\cos\phi + \op{a}_{2}\sin\phi\right) + \omega_{-}\op{a}_{1}^{\dagger}\op{a}_{1}\\ 
    &+ \omega_{+}\left(\op{a}_{2}^{\dagger}\cos\phi - \op{a}_{3}^{\dagger}\sin\phi\right)\left(\op{a}_{2}\cos\phi - \op{a}_{3}\sin\phi\right)\\ 
    &+ \sqrt{2} g\left[\left(\op{a}_{3}\cos\phi + \op{a}_{2}\sin\phi\right)\left(\op{a}_{2}^{\dagger}\cos\phi - \op{a}_{3}^{\dagger}\sin\phi\right)\right.\\
    &+ \left.\left(\op{a}_{3}^{\dagger}\cos\phi + \op{a}_{2}^{\dagger}\sin\phi\right)\left(\op{a}_{2}\cos\phi-\op{a}_{3}\sin\phi\right)\right],
\end{align*}
that simplifies to
\begin{equation}\label{H2}
\begin{aligned}
    \op{H}_{2} &= \op{a}_{3}^{\dagger}\op{a}_{3}(\omega\cos^{2}\phi+\omega_{+}\sin^{2}\phi)+\op{a}_{2}^{\dagger}\op{a}_{2}(\omega\sin^{2}\phi+\omega_{+}\cos^{2}\phi)\\
    &+\omega_{-}\op{a}_{1}^{\dagger}\op{a}_{1}-\sqrt{2}g(\op{a}_{3}\op{a}_{3}^{\dagger}-\op{a}_{2}\op{a}_{2}^{\dagger})\sin2\phi\\
    &+\frac{1}{2}(\op{a}_{3}^{\dagger}\op{a}_{2}+\op{a}_{2}^{\dagger}\op{a}_{3})\left([\omega-\omega_{+}]\sin2\phi+2\sqrt{2}g\cos2\phi\right),
\end{aligned}
\end{equation}
where we can dismiss the last interacting term, by setting adequately the angle $\phi$, for this we have to solve the equation
\begin{equation*}
    (\omega-\omega_{+})\sin2\phi+2\sqrt{2}g\cos2\phi=0,
\end{equation*}
such that (twice) the angle is given by
\begin{equation*}
    2\phi = \arctan\left(\frac{2\sqrt{2}g}{\omega_{+}-\omega}\right).
\end{equation*}

Then, the Hamiltonian in equation \eqref{H2} may be simplified  as
\begin{equation}\label{H2r}
\begin{aligned}
    H_{2} &= \op{a}_{3}^{\dagger}\op{a}_{3}(\omega\cos^{2}\phi + \omega_{+}\sin^{2}\phi - \sqrt{2}g\sin2\phi)\\
    &+ \op{a}_{2}^{\dagger}\op{a}_{2}(\omega\sin^{2}\phi + \omega_{+}\cos^{2}\phi + \sqrt{2}g\sin2\phi) + \omega_{-}\op{a}_{1}^{\dagger}\op{a}_{1}
\end{aligned}
\end{equation}
where we terms inside the parenthesis can be reduced to the compact form
\begin{align*}
    &\omega\cos^{2}\phi+\omega_{+}\sin^{2}\phi-\sqrt{2}g\sin2\phi\\
    &= \frac{\omega}{2}\left(1+\cos2\phi\right)+\frac{\omega_{+}}{2}\left(1-\cos2\phi\right)-\sqrt{2}g\sin2\phi\\
    &= \frac{1}{2}\left(\omega-\omega_{+}\right)\cos2\phi + \frac{1}{2}\left(\omega_{+} + \omega\right) - \sqrt{2}g\sin2\phi\\
    &= \frac{1}{2}\left(\omega_{+} + \omega\right) - \frac{1}{2}\frac{\left(\omega_{+} - \omega\right)^{2}}{\sqrt{8g^{2}+\left(\omega_{+} - \omega\right)^{2}}} - \frac{4g^{2}}{\sqrt{8g^{2}+\left(\omega_{+} - \omega\right)^{2}}}\\
    &= \frac{1}{2}\left(\omega_{+} + \omega\right) - \frac{1}{2}\left(\frac{8g^{2}+\left(\omega_{+} - \omega\right)^{2}}{\sqrt{8g^{2}+\left(\omega_{+} - \omega\right)^{2}}}\right)\\
    &= \frac{1}{2}\left(\left[\omega_{+} + \omega\right] - \sqrt{8g^{2}+\left(\omega_{+} - \omega\right)^{2}}\right) = \Omega,
\end{align*}
and samewise, for the other term, we obtain
\begin{align*}
    \omega\sin^{2}\phi+\omega_{+}\cos^{2}\phi + \sqrt{2}g\sin2\phi &=
     \frac{1}{2}\left(\left[\omega_{+} + \omega\right] +\sqrt{8g^{2}+\left(\omega_{+} - \omega\right)^{2}}\right)\\ &= \Omega_{2}.
\end{align*}
Finally, we arrive to the diagonal Hamiltonian for the uncoupled harmonic oscillators
\begin{equation}\label{H2D}
    \op{H}_{2} = \Omega_{2} \op{a}^{\dagger}_{2}\op{a}_{2} + \omega_{-}\op{a}^{\dagger}_{1}\op{a}_{1} + \Omega \op{a}_{3}^{\dagger}\op{a}_{3},
\end{equation}
where the effective frequencies $\Omega$, $\Omega_{2}$, and $\omega_{-}$, are the eigenvalues that diagonalize the original Hamiltonian (\ref{H3}).

\subsection{Full solution}\label{sec:fulls}
\subsubsection*{Transformed initial condition}
With the diagonal Hamiltonian \eqref{H2D}, we can now easily factorize the exponential \eqref{expt} as
\begin{equation}\label{Expt}
    e^{-\mi k\frac{\op{H}_{2}}{\gamma}} = e^{-\frac{\mi k}{\gamma}\omega_{-} \op{a}^{\dagger}_{1}\op{a}_{1}}e^{-\frac{\mi k}{\gamma}\Omega_{2} \op{a}^{\dagger}_{2}\op{a}_{2}}e^{-\frac{\mi k}{\gamma}\Omega \op{a}_{3}^{\dagger}\op{a}_{3}}
\end{equation}
that may be easily applied to an initial wavefunction $\ket{\psi(0)}$. For the $k$-th element we have
\begin{equation}\label{pk}
    \ket{\psi_{k}}=\op{R}_{12}^{\dagger}\op{R}_{2}^{\dagger}\,e^{-\mi k\frac{\op{H}_2}{\gamma}}\op{R}_{2}\op{R}_{12}\ket{\psi(0)}.
\end{equation}
Next we show how to apply the set of operators in the above equation to a particular initial wave function. We choose one of the oscillators to be given in a coherent state while the other two are given in their vacuum states, namely
\begin{equation*}
\ket{\psi(0)} = \ket{\alpha}_{1}\ket{0}_{2}\ket{0}_{3} = e^{\alpha \op{a}_{1}^{\dagger}-\alpha^{*}\op{a}_{1}}\ket{0}_{1}\ket{0}_{2}\ket{0}_{3},
\end{equation*}
yielding the $k$-th element of $\ket{\psi_k}$ 
\begin{equation*}
    \ket{\psi_{k}}=\op{R}_{12}^{\dagger}\op{R}_{2}^{\dagger}\,e^{-\mi k\frac{\op{H}_{2}}{\gamma}}\op{R}_{2}\op{R}_{12}e^{\alpha \op{a}_{1}^{\dagger}-\alpha^{*}\op{a}_{1}}\ket{0}_{1}\ket{0}_{2}\ket{0}_{3},
\end{equation*}
that, by properly applying a unit operator may be rewritten as
\begin{equation*}
    \ket{\psi_{k}}=\op{R}_{12}^{\dagger}\op{R}_{2}^{\dagger}e^{-\mi k\frac{\op{H}_2}{\gamma}}\op{R}_{2}\op{R}_{12}\,e^{\alpha \op{a}_{1}^{\dagger}-\alpha^{*}\op{a}_{1}}\op{R}_{12}^{\dagger}\op{R}_{2}^{\dagger}\op{R}_{2}\op{R}_{12}\ket{0}_{1}\ket{0}_{2}\ket{0}_{3},
\end{equation*}
and because the initial condition is invariant under the action of $\op{R}_{2}\op{R}_{12}$, we arrive to
\begin{equation}\label{initcond}
    \ket{\psi_{k}}=\op{R}_{12}^{\dagger}\op{R}_{2}^{\dagger}\,e^{-\mi k\frac{\op{H}_2}{\gamma}}\op{R}_{2}\op{R}_{12}\,e^{\alpha \op{a}_{1}^{\dagger}-\alpha^{*}\op{a}_{1}}\op{R}_{12}^{\dagger}\op{R}_{2}^{\dagger}\ket{0}_{1}\ket{0}_{2}\ket{0}_{3}.
\end{equation}
Explicit expressions for the action of the transformations on the displacement operator are not difficult to achieve, in fact we may obtain
\begin{equation}\label{dops}
\begin{aligned}
    &\op{R}_{2}\op{R}_{12}e^{\alpha\op{a}_{1}^{\dagger} - \alpha^{*} \op{a}_{1}} \op{R}_{12}^{\dagger}\op{R}_{2}^{\dagger}\\ &= e^{\op{R}_{2}\op{R}_{12}\left(\alpha \op{a}_{1}^{\dagger} - \alpha^{*} \op{a}_{1}\right)\op{R}_{12}^{\dagger}\op{R}_{2}^{\dagger}}\\
    &= e^{\op{R}_{2}\left(\left[\alpha \op{a}_{1}^{\dagger} - \alpha^{*}\op{a}_{1}\right]\cos\theta +\left[\alpha \op{a}_{2}^{\dagger} - \alpha^{*}\op{a}_{2}\right]\sin\theta \right)\op{R}_{2}^{\dagger}}\\
    &= e^{\left(\alpha \op{a}_{1}^{\dagger} - \alpha^{*}\op{a}_{1}\right)\cos\theta + \left(\alpha\left[\op{a}_{2}^{\dagger}\cos\phi - \op{a}_{3}^{\dagger}\sin\phi\right] - \alpha^{*}\left[\op{a}_{2}\cos\phi - \op{a}_{3}\sin\phi\right]\right)\sin\theta}\\
    &= e^{\left(\alpha \op{a}_{1}^{\dagger} - \alpha^{*} \op{a}_{1}\right)\cos\theta}\times e^{\left(\alpha \op{a}_{2}^{\dagger} - \alpha^{*} \op{a}_{2}\right)\cos\phi\sin\theta}\times e^{-\left(\alpha \op{a}_{3}^{\dagger} - \alpha^{*} \op{a}_{3}\right)\sin\phi\sin\theta},
\end{aligned}
\end{equation}
such that factorization of the initial displacement operator may be obtained via the transformed displacement operators that now depend on the angles $\theta=\pi/4$, and $\phi$ previously obtained.

The task now is to determine the explicit action of the displacement-like operators \eqref{dops} on the initial condition kets. Because they are now factorized, their action affect only the kets with the same subscripts (or lack of it) to give us
\begin{equation}\label{ticond}
\begin{aligned}
    &e^{\left(\alpha \op{a}_{1}^{\dagger} - \alpha^{*} \op{a}_{1}\right)\cos\theta}\times e^{\left(\alpha \op{a}_{2}^{\dagger} - \alpha^{*} \op{a}_{2}\right)\cos\phi\sin\theta}\times e^{-\left(\alpha \op{a}_{3}^{\dagger} - \alpha^{*} \op{a}_{3}\right)\sin\phi\sin\theta}\ket{0}_{1}\ket{0}_{2}\ket{0}_{3} =\\
    &\qquad \ket{\alpha\cos\theta}_{1}\ket{\alpha\cos\phi\sin\theta}_{2}\ket{-\alpha\sin\phi\sin\theta}_{3}.
\end{aligned}
\end{equation}

Finally, we can observe the action of the Hamiltonian exponential operator, and the dagger operators of \eqref{initcond} over \eqref{ticond}, such that we may obtain the $k$-the element as
 \begin{align*}
    \ket{\psi_{k}} &= \op{R}_{12}^{\dagger}\op{R}_{2}^{\dagger}\,e^{-\mi k\frac{\op{H}_{2}}{\gamma}}\ket{\alpha\cos\theta}_{1}\ket{\alpha\cos\phi\sin\theta}_{2}\ket{-\alpha\sin\phi\sin\theta}_{3}\\
    &= \op{R}_{12}^{\dagger}\op{R}_{2}^{\dagger}\ket{e^{-\frac{\mi k}{\gamma}\Omega_{2}}\alpha\cos\theta}_{1}\ket{e^{-\frac{\mi k}{\gamma}\omega_{-}}\alpha\cos\phi\sin\theta}_{2}\ket{-e^{-\frac{\mi k}{\gamma}\Omega}\alpha\sin\phi\sin\theta}_{3}.
\end{align*}

\subsubsection*{Oscillator modes expectation values}
To calculate the average number of photons for each mode,  $\braket{\op{a}_{j}^{\dagger}\op{a}_{j}}$, we use equation \eqref{average}. For that we need to evaluate the expression 
\begin{equation*}
    \braket{\psi_{k}|\op{a}_{j}^{\dagger}\op{a}_{j}|\psi_{k}},
\end{equation*}
for each oscillator. Starting with the oscillator related to the number operator $\op{a}_{3}^{\dagger}\op{a}_{3}$ we have
\begin{equation}\label{photons0}
\begin{aligned}
    \braket{\psi_{k}|\op{a}_{3}^{\dagger}\op{a}_{3}|\psi_{k}} = &\bra{-e^{\frac{\mi k}{\gamma}\Omega}\alpha\sin\phi\sin\theta}_{3}\bra{e^{\frac{\mi k}{\gamma}\Omega_{2}}\alpha\cos\phi\sin\theta}_{2}\bra{e^{\frac{\mi k}{\gamma}\omega_{-}}\alpha\cos\theta}_{1}\\ &\times \op{R}_{2}\op{R}_{12}\op{a}^{\dagger}\op{a}  \op{R}_{12}^{\dagger}\op{R}_{2}^{\dagger}\ket{e^{-\frac{\mi k}{\gamma}\omega_{-}}\alpha\cos\theta}_{1}\\ &\hspace{1cm}\times \ket{e^{-\frac{\mi k}{\gamma}\Omega_{2}}\alpha\cos\phi\sin\theta}_{2}\ket{-e^{-\frac{\mi k}{\gamma}\Omega}\alpha\sin\phi\sin\theta}_{3}
\end{aligned}
\end{equation}
and the similarity transformation carried by the operators $\op{R}_{2}\op{R}_{12}$ gives
\begin{equation}\label{ph00r}
    \op{R}_{2}\op{R}_{12}\op{a}_{3}^{\dagger}\op{a}_{3}  \op{R}_{12}^{\dagger}\op{R}_{2}^{\dagger}=\op{a}_{3}^{\dagger}\op{a}_{3}\cos^2\phi+\op{a}_2^{\dagger}\op{a}_2\sin^2\phi+(\op{a}_{3}^{\dagger}\op{a}_{2}+\op{a}_2^{\dagger}\op{a}_{3})\cos\phi\sin\phi.
\end{equation}

After this, the calculation of equation \eqref{photons0} may be easily carried out by applying the annihilation operators to the coherent states to the right and the creation operators to the coherent states to the left. The sum given in \eqref{average} may be added to obtain explicit expressions.
For instance, we explicitly obtain the expectation value
\begin{equation}\label{photons00}
    \braket{\psi_{k}|\op{a}_{3}^{\dagger}\op{a}_{3}|\psi_{k}}={|\alpha|^2}\cos^2\phi\sin^2 \phi\left(1-\cos\left[\frac{k}{\gamma}(\Omega-\Omega_{2})\right]\right),
\end{equation}
that can be inserted in equation \eqref{average}, and yields the average number of photons for the mode 3
\begin{equation}\label{average00}
    \braket{a_{3}^{\dagger}a_{3}} = {|\alpha|^2}\cos^2\phi\sin^2 \phi \left(1-\frac{e^{-\gamma t}}{2}\left[e^{\gamma t e^{\mi\frac{\Omega-\Omega_2}{\gamma}}}+e^{\gamma t e^{-\mi\frac{\Omega-\Omega_2}{\gamma}}}\right]\right).
\end{equation}

Analogously for the  other modes, expressions for their average number of excitations,  $\braket{\op{a}_{1}^{\dagger}\op{a}_{1}}$ and $\braket{\op{a}_{2}^{\dagger}\op{a}_{2}}$, may be obtained explicitly. We  follow the same procedure as above and  calculate the action of the $\op{R}$'s operators, for mode $1$
\begin{align*}
    \op{R}_{2}\op{R}_{12}\,\op{a}_{1}^{\dagger}\op{a}_{1}\,\op{R}_{12}^{\dagger}\op{R}_{2}^{\dagger} &= \frac{1}{2}\left[\op{a}_{1}^{\dagger}\op{a}_{1} + \op{a}_{2}^{\dagger}\op{a}_{2}\cos^{2}\phi + \op{a}_{3}^{\dagger}\op{a}_{3}\sin^{2}\phi\right.\\ &\left.-\left(\op{a}_{2}^{\dagger}\op{a}_{3} + \op{a}_{3}^{\dagger}\op{a}_{2}\right)\sin\phi\cos\phi
    +\left(\op{a}_{2}^{\dagger}\op{a}_{1} + \op{a}_{2}^{\dagger}\op{a}_{1}\right)\cos\phi\right.\\ &\left.-\left(\op{a}_{1}^{\dagger}\op{a}_{3} + \op{a}_{3}^{\dagger}\op{a}_{1}\right)\sin\phi\right],
\end{align*}
and mode $2$
\begin{align*}
    \op{R}_{2}\op{R}_{12}\,\op{a}_{2}^{\dagger}\op{a}_{2}\,\op{R}_{12}^{\dagger}\op{R}_{2}^{\dagger} &= \frac{1}{2}\left[\op{a}_{1}^{\dagger}\op{a}_{1} + \op{a}_{2}^{\dagger}\op{a}_{2}\cos^{2}\phi + \op{a}_{3}^{\dagger}\op{a}_{3}\sin^{2}\phi\right.\\ &\left.-\left(\op{a}_{2}^{\dagger}\op{a}_{3} + \op{a}_{3}^{\dagger}\op{a}_{2}\right)\sin\phi\cos\phi
    -\left(\op{a}_{2}^{\dagger}\op{a}_{1} + \op{a}_{2}^{\dagger}\op{a}_{1}\right)\cos\phi\right.\\ &\left.+\left(\op{a}_{1}^{\dagger}\op{a}_{3} + \op{a}_{3}^{\dagger}\op{a}_{1}\right)\sin\phi\right],
\end{align*}
for which we calculate the expectation values for the $k$-th element, for modes 1 and 2, respectively as
\begin{align*}
    &\bra{\psi_{k}}\op{a}_{1}^{\dagger}\op{a}_{1}\ket{\psi_{k}}\\ 
    &\quad= \frac{1}{4}|\alpha|^{2}\left[1 + \left(\cos^{4}\phi + \sin^{4}\phi\right) + 2\sin^{2}\phi\cos^{2}\phi\cos\left(\frac{k}{\gamma}\left(\Omega-\Omega_{2}\right)\right)\right.\\
    &\qquad + \left.2\cos^{2}\phi\cos\left(\frac{k}{\gamma}\left(\omega_{-}-\Omega_{2}\right)\right)+ 2\sin^{2}\phi\cos\left(\frac{k}{\gamma}\left(\omega_{-}-\Omega\right)\right)\right]
\end{align*}
and
\begin{align*}
    &\bra{\psi_{k}}\op{a}_{2}^{\dagger}\op{a}_{2}\ket{\psi_{k}}\\
    &\quad= \frac{1}{4}|\alpha|^{2}\left[1 + \left(\cos^{4}\phi + \sin^{4}\phi\right) + 2\sin^{2}\phi\cos^{2}\phi\cos\left(\frac{k}{\gamma}\left(\Omega-\Omega_{2}\right)\right)\right.\\
    &\qquad - \left.2\cos^{2}\phi\cos\left(\frac{k}{\gamma}\left(\omega_{-}-\Omega_{2}\right)\right)- 2\sin^{2}\phi\cos\left(\frac{k}{\gamma}\left(\omega_{-}-\Omega\right)\right)\right].
\end{align*}

\begin{figure}[htbp]
    \centering
    \includegraphics[width = \textwidth, height = 0.8\textheight, keepaspectratio]{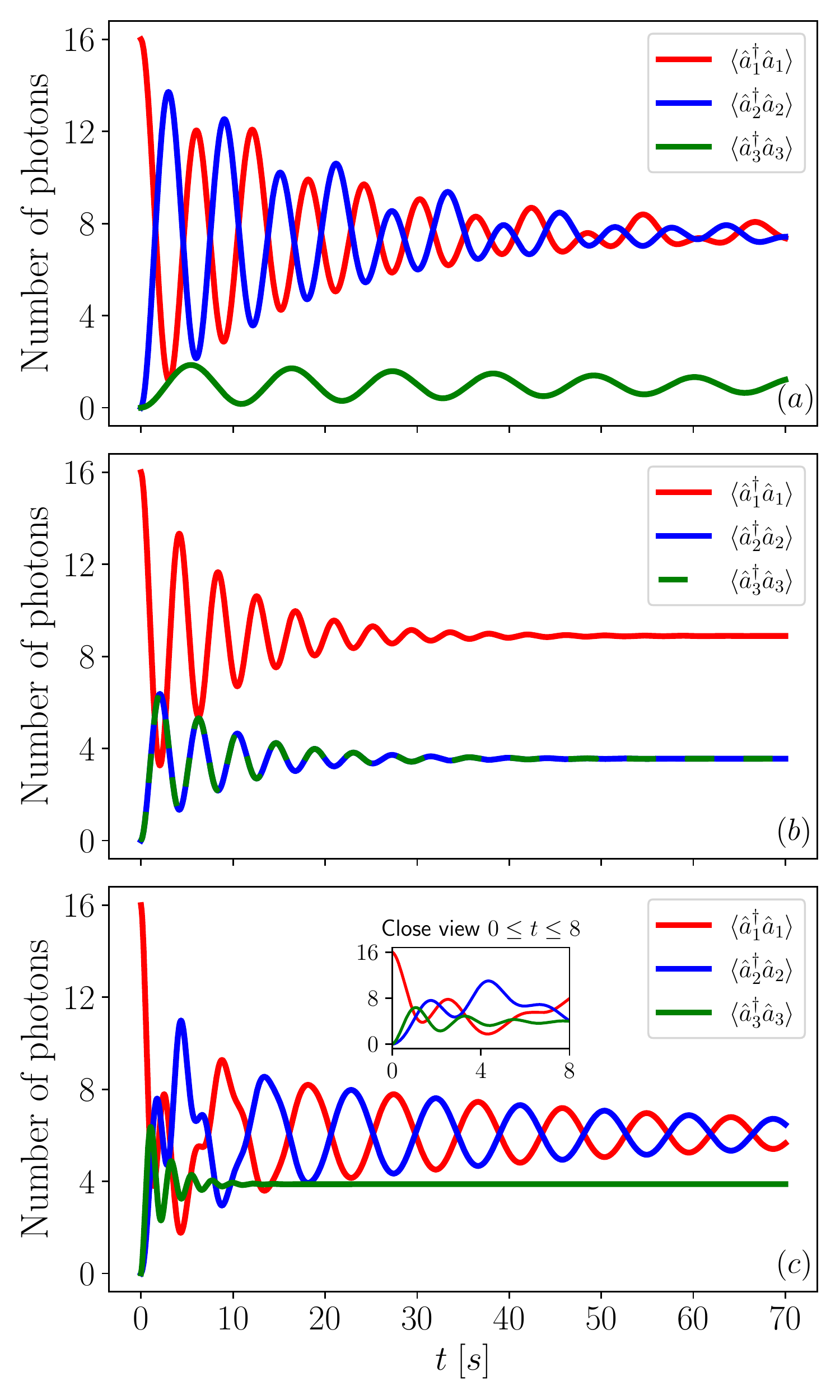}
    \caption{Plot showing the photon number oscillations given by equations \eqref{average00}, \eqref{average11}, and \eqref{average22}. We have the parameters: $\omega = 4$, $\lambda = 0.5$, $\gamma = 10$, $\alpha=4$. We decide to variate the parameter $g$, who gives the coupling strength between the three oscillator, then we have in descending order (a) $g=0.1$, (b) $g=0.5$, where oscillators 2 and 3 overlaps, and (c) $g=1$.}
    \label{fig:expect1}
\end{figure}

\begin{figure}[htbp]
    \centering
    \includegraphics[width = \textwidth, height = 0.8\textheight, keepaspectratio]{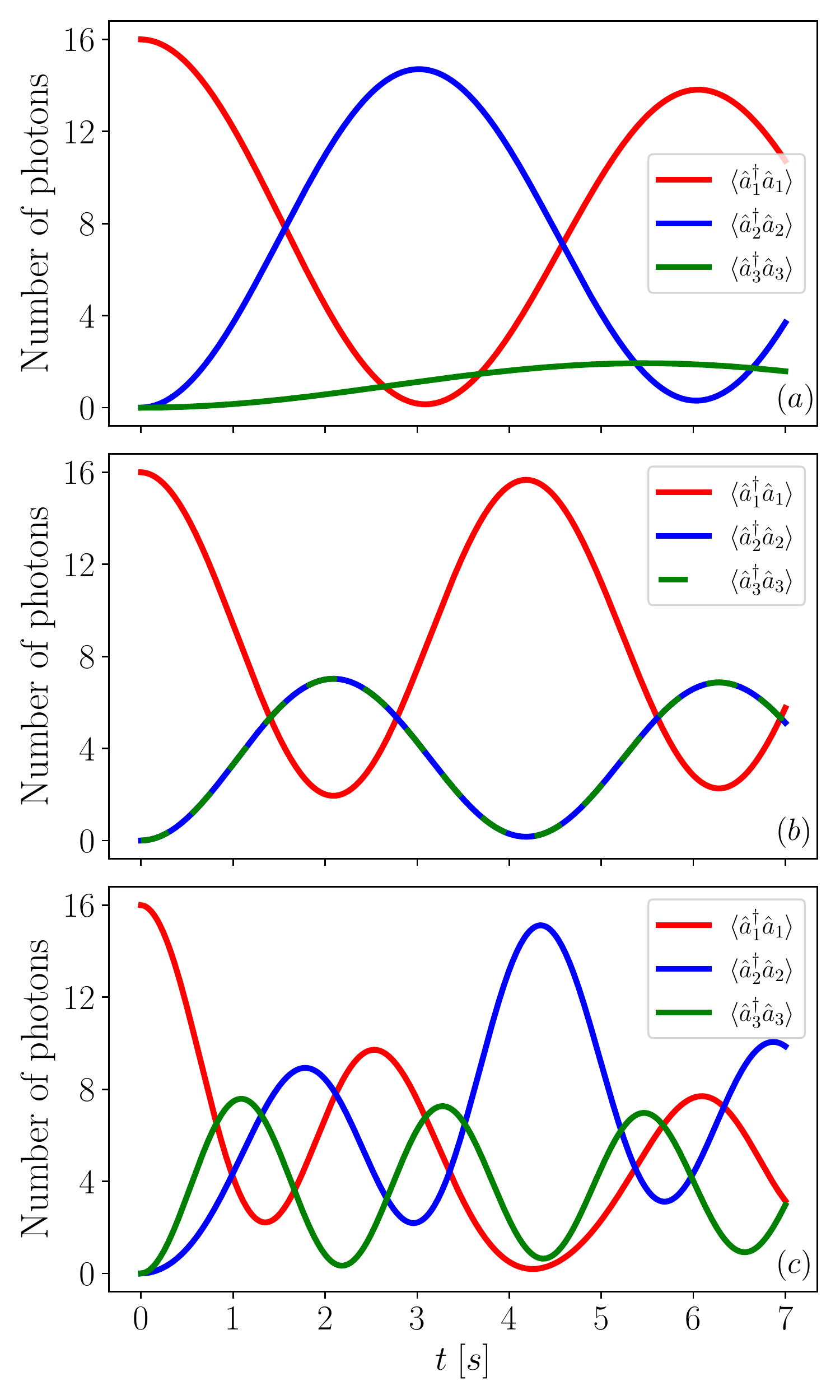}
    \caption{Plot showing the photon number oscillations given by equations \eqref{average00}, \eqref{average11}, and \eqref{average22}. We have the parameters: $\omega = 4$, $\lambda = 0.5$, $\gamma = 100$, $\alpha=4$. We decide to variate the parameter $g$, who gives the coupling strength between the three oscillator, then we have in descending order (a) $g=0.1$, (b) $g=0.5$, where oscillators 2 and 3 overlaps, and (c) $g=1$.}
    \label{fig:expect2}
\end{figure}

With this last couple of equations, we can finally produce the photon number evolution for oscillator 1 and 2, that are explicitly given by
\begin{equation}\label{average11}
\begin{aligned}
    &\braket{\op{a}_{1}^{\dagger}{\op{a}_{1}}} = \frac{1}{4}|\alpha|^{2}\times\\
    & \left(1 + \frac{1}{4}\left(3+\cos\left(4\phi\right)\right) + e^{-\gamma t}\left[\left\{e^{\gamma t e^{i\frac{\Omega-\Omega_{2}}{\gamma}}}+e^{\gamma t e^{-i\frac{\Omega-\Omega_{2}}{\gamma}}}\right\}\cos^{2}\phi\sin^{2}\phi\right.\right. +\\
    & \left.\left.\left\{e^{\gamma t e^{i\frac{\omega_{-}-\Omega_{2}}{\gamma}}}+e^{\gamma t e^{-i\frac{\omega_{-}-\Omega_{2}}{\gamma}}}\right\}\cos^{2}\phi + \left\{e^{\gamma t e^{i\frac{\omega_{-}-\Omega}{\gamma}}}+e^{\gamma t e^{-i\frac{\omega_{-}-\Omega}{\gamma}}}\right\}\sin^{2}\phi\right]\right),
\end{aligned}
\end{equation}

\begin{equation}\label{average22}
\begin{aligned}
    &\braket{\op{a}_{2}^{\dagger}{\op{a}_{2}}} = \frac{1}{4}|\alpha|^{2}\times\\
    &\left(1 + \frac{1}{4}\left(3+\cos\left(4\phi\right)\right) + e^{-\gamma t}\left[\left\{e^{\gamma t e^{i\frac{\Omega-\Omega_{2}}{\gamma}}}+e^{\gamma t e^{-i\frac{\Omega-\Omega_{2}}{\gamma}}}\right\}\cos^{2}\phi\sin^{2}\phi\right.\right.-\\
    & \left.\left.\left\{e^{\gamma t e^{i\frac{\omega_{-}-\Omega_{2}}{\gamma}}}+e^{\gamma t e^{-i\frac{\omega_{-}-\Omega_{2}}{\gamma}}}\right\}\cos^{2}\phi - \left\{e^{\gamma t e^{i\frac{\omega_{-}-\Omega}{\gamma}}}+e^{\gamma t e^{-i\frac{\omega_{-}-\Omega}{\gamma}}}\right\}\sin^{2}\phi\right]\right).
\end{aligned}
\end{equation}

\newpage
\subsection{Results}\label{sec:results}
In Figures \ref{fig:expect1} and \ref{fig:expect2},  we plot equations \eqref{average00}, \eqref{average11}, and \eqref{average22}, for a set of parameters $\{\omega = 4$, $\lambda = 0.5$, $\gamma = 10$, $\alpha=4\}$, and $\{\omega = 4$, $\lambda = 0.5$, $\gamma = 100$, $\alpha=4\}$. In both cases, the increasing parameter is the coupling strength $g = \{0.1, 0.5, 1\}$, such that we can see how the energy is exchanged between the modes and intrinsic decoherence produces a damping of the oscillations which is slower as we increase parameter $\gamma$. Fig. \ref{fig:expect1} shows that as $g$ increases the oscillations for the mode $a$ damp faster as its interaction to the other two modes increase. Oscillations in the red curve may be seen to stop faster although the same intrinsic decoherence rate is set for that figure. In Fig. \ref{fig:expect2}, as we decrease the intrinsic decoherence rate, {\it i.e.}, increase the value of $\gamma$ the oscillations are maintained for greater interaction times.

A reproducible software code has been developed to support the numerical findings presented here \cite{alejandro_r_urzua_2021_5131447}. %There you can manipulate equations \eqref{average00}, \eqref{average11}, and \eqref{0020}; also you can compare with a numerically solved traditional Schr\"odinger equation, who doesn't show decoherence features.

\section{Conclusions}\label{sec:concls}
We have given a complete solution to the Milburn equation and shown that the average of arbitrary operators may be easily calculated. We have applied the solution to study  intrinsic decoherence  in the interaction of three quantized fields under the rotating wave approximation.

\printbibliography

\end{document}